\documentclass{article}

\usepackage{arxiv}

\usepackage[utf8]{inputenc} 
\usepackage[T1]{fontenc}    
\usepackage{booktabs}       
\usepackage{amsfonts}       
\usepackage{nicefrac}       
\usepackage{microtype}      
\usepackage{lipsum}		
\usepackage{graphicx}
\usepackage{natbib}
\usepackage{doi}

\usepackage{scalerel}
\usepackage{amsmath}
\usepackage{mathtools} 
\usepackage{bm}
\usepackage{float}
\usepackage{scalerel}

\usepackage{RJournal}
\usepackage{amsmath,amssymb,array}
\usepackage{booktabs}

\title{BayesDLMfMRI: Bayesian Matrix-Variate Dynamic Linear Models for Task-based fRMI Modeling in \textbf{R}}

\author{ \href{https://orcid.org/0000-0002-6370-8837}{\includegraphics[scale=0.06]{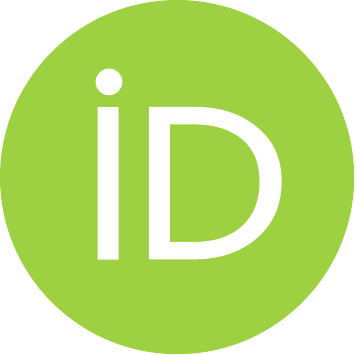}\hspace{1mm}Johnatan Cardona~Jiménez}\thanks{Use footnote for providing further
		information about author (webpage, alternative
		address)---\emph{not} for acknowledging funding agencies.} \\
	Facultad de Ingeniería\\
	Institución Universitaria Pascual Bravo\\
	Cl 73 No. 73A - 226, Medellín, ZIP 050034, Colombia \\
	\texttt{johnatan.cardona@pascualbravo.edu.co} \\
}

\renewcommand{\headeright}{ }
\renewcommand{\undertitle}{Technical Report}
\renewcommand{\shorttitle}{BayesDLMfMRI: Bayesian Matrix-Variate Dynamic Linear Models for Task-based fRMI Modeling in \textbf{R}}

\hypersetup{
pdftitle={A template for the arxiv style},
pdfsubject={q-bio.NC, q-bio.QM},
pdfauthor={David S.~Hippocampus, Elias D.~Striatum},
pdfkeywords={First keyword, Second keyword, More},
}

\begin{document}
\maketitle

\begin{abstract}
	This article introduces an R package to perform statistical analysis for task-based fMRI data at both individual and group levels. The analysis to detect brain activation at the individual level is based on modeling the fMRI signal using Matrix-Variate Dynamic Linear Models (MDLM). Therefore, the analysis for the group stage is based on posterior distributions of the state parameter obtained from the modeling at the individual level. In this way, this package offers several R functions with different algorithms to perform inference on the state parameter to assess brain activation for both individual and group stages. Those functions allow for parallel computation when the analysis is performed for the entire brain as well as analysis at specific voxels when it is required.
\end{abstract}

\keywords{First keyword \and Second keyword \and More}

\section{Introduction: fMRI packages with options for Bayesian Analysis}

Statistical modeling of processed fMRI data is a challenging problem, which has caught the attention of the statistical community in the past two decades. Large amount of observations are usually obtained with only one subject in an fMRI session which makes the implementation of more appropriate and sophisticated statistical models that can account for the spatiotemporal structures that are usually present in this type of data really challenging. From a Bayesian perspective, there have been published and proposed different types of models to model fMRI data (see for instance \citet{zhang2016spatiotemporal}, \citet{eklund2017bayesian}, \citet{bezener2018bayesian}, \citet{yu2018bayesian} and a complete review by \citet{zhang2015bayesian}). Despite that, just a few software implementations of Bayesian methods are available for fMRI data analysis. For instance, the \pkg{FSL} software \citep{jenkinson2012fsl} offers a Bayesian procedure for the group stage analysis, which depends on a Frequentist or Classical output from the individual stage. Another popular software among practioners in the neurosicience community with implementations of Bayesian methods is the \pkg{SPM} package \citep{penny2011statistical}. It has alternatives for Bayesian analysis to both individual and group stages, but as in the case of the \pkg{FSL} package, the group analysis depends on a Frequentist individaul stage output. There is also available a \pkg{MATLAB GUI} called \pkg{NPBayes-fMRI} \citep{kook2019npbayes}, which is an implementation of the work proposed by \citet{zhang2016spatiotemporal}, a fully Bayesian modeling for individual and group stages. In R \citep{Rcite}, tools for different types of fMRI data analysis are provided in packages like \pkg{fmri} \citep{fmriPackage}, \pkg{oro.nifti} \citep{oro.nifti}, \pkg{neurobase} \citep{neurobase}, and \pkg{neuRosim} \citep{welvaert2011neurosim}. For fMRI data analysis under a Bayesian approach there was a package called \pkg{cudaBayesreg} \citep{da2011cudabayesreg}, which was removed recently from the CRAN repository. Thus, the package \pkg{BHMSMAfMRI} \citep{BHMSMAfMRI} is to our knowledge the only Bayesian option for fMRI data analysis in R to this date.\\ \\
In this work, we are presenting an R package called \pkg{BayesDLMfMRI}, which is an implementation of the method proposed in \citet{jimnez2019assessing} and \citet{jimnez2019assessing2}. The \pkg{BayesDLMfMRI} permits to perform fMRI individual and group analysis based on the Matrix-Variate Dynamic Linear Model (MVDLM) proposed by \citet{quintana1985}. Given this type of analysis usually involved large amounts of data, we take advantage of the packages  \pkg{Rcpp} \citep{eddelbuettel2011rcpp} and \pkg{RcppArmadillo} \citep{RcppArma} in order to speed up the computation time. \pkg{BayesDLMfMRI} also depends internally on the package \pkg{pbapply} \citep{pbapplyref}, which allows the user to visualize a progress bar while the process is run in either sequence or parallel. In order to run an analysis using our package the user must provide a 4D array (or a 4D$\times N$ array for a group analysis with $N$ subjects) containing the sequence of processed images and a design matrix whose columns are related to the so-called expected blood-oxygen-Level dependent  (\textbf{BOLD}) response and/or some other covariates related to particular subjects' characteristics. To process the raw images, we recommend the use of packages such as \pkg{FSL} or \pkg{SPM}. And to build the expected BOLD response the user has options like the \pkg{fmri} package, which allows defining different types of models to represent the haemodynamic response function (\textbf{HRF}). \pkg{BayesDLMfMRI} package is intended to be just another well tested and assessed tool for the practitioners who are interested to perform statistical analysis of processed fMRI images from a Bayesian perspective. In the next section, we give a brief description of the model and methods on which the \pkg{BayesDLMfMRI} package is based and present its R functions to perform fMRI data analysis. In section three, we present some examples to illustrate the use of the package and in the final section, we give some concluding remarks.

\section{Methods and software}

\begin{table}[H]
\caption{\label{tab:1} Functions implemented in the \pkg{BayesDLMfMRI} package for fMRI (group) individual analysis. LLT stands for Transformation Linear Test, which is the same as the average cluster effect (ACE) distribution defined in ArchiveRef.}
\centering
\begin{tabular}{lp{6.4cm}}
\hline
Function              & Description \\ \hline
(\code{ffdGroupEvidenceFEST})      \code{ffdEvidenceFEST}  & It returns 3D arrays to build (group) individual activation evidence maps  (based on outputs from MVDLM) fitting an MVDLM at individual level and using the FEST algorithm to assess voxel activation. There are two options related to the posterior distribution of $\Theta_{\scaleto{[i,j,k]t\mathstrut}{4pt}}^{\scaleto{(z)\mathstrut}{4pt}}$: LLT and Joint. Each one runs independently and must be set by the user as an input parameter.\\
(\code{ffdGroupEvidenceFFBS}) \code{ffdEvidenceFFBS}      & It returns 3D arrays to build (group) individual activation evidence maps  (based on outputs from MVDLM) fitting an MVDLM and using the FFBS algorithm to assess voxel activation. The options LLT and Joint related to the posterior distribution of $\Theta_{\scaleto{[i,j,k]t\mathstrut}{4pt}}^{\scaleto{(z)\mathstrut}{4pt}}$ are simultaneously executed in the same run.\\
(\code{ffdGroupEvidenceFSTS}) \code{ffdEvidenceFSTS}    &  Same features as (\code{ffdGroupEvidenceFFBS}) \code{ffdEvidenceFFBS}, though using the FSTS algorithm.       \\
(\code{GroupSingleVoxelFEST}) \code{SingleVoxelFEST} & Produces some usefull outputs from a single voxel analysis related to the FEST algorithm \\   
(\code{GroupSingleVoxelFFBS}) \code{SingleVoxelFFBS}  & Same features as (\code{GroupSingleVoxelFEST}) \code{SingleVoxelFEST}, though using the FFBS algorithm\\
(\code{GroupSingleVoxelFSTS}) \code{SingleVoxelFSTS}  & Same features as (\code{GroupSingleVoxelFEST}) \code{SingleVoxelFEST}, though using the FSTS algorithm   \\ 
   \hline
\end{tabular}
\end{table}

The type of MVDLM which this package relies on is the version originally developed by \citet{quintana1985} and \citet{quintana1987multivariate}. Here, we just give a brief description of the model and methods. For a better understanding of the method implemented in this package for the individual stage, see \citet{jimnez2019assessing}. Let $\mathbf{Y}_{\scaleto{[i,j,k]t\mathstrut}{6pt}}^{(z)}$ be a $q \times 1$ random vector representing the cluster of observed BOLD responses at position $(i,j,k)$ in the brain image, time $t$ and subject $z$, for $i=1, \ldots, d_1$, $j=1,\ldots,d_2$, $k=1,\ldots,d_3$, $t=1,\ldots, T$ and $z=1\ldots,N$. Thus, the cluster of BOLD signals is modeled as

\begin{equation} \label{sec2:equ1}
\begin{array}{lccl}
\text{Observation:}\ &\mathbf{Y}_{\scaleto{[i,j,k]t\mathstrut}{4pt}}^{\scaleto{(z)\mathstrut}{4pt}} & =& \mathbf{F}^{'}_t\Theta_{\scaleto{[i,j,k]t\mathstrut}{4pt}}^{\scaleto{(z)\mathstrut}{4pt}} + \bm{\nu}_{\scaleto{[i,j,k]t\mathstrut}{4pt}}^{\scaleto{'(z)\mathstrut}{4pt}} \\
 \text{Evolution:}\ &\Theta_{\scaleto{[i,j,k]t\mathstrut}{4pt}}^{\scaleto{(z)\mathstrut}{4pt}}& =& \mathbf{G}_t\Theta_{\scaleto{[i,j,k]t-1\mathstrut}{4pt}}^{\scaleto{(z)\mathstrut}{4pt}} + \Omega_{\scaleto{[i,j,k]t\mathstrut}{4pt}}^{\scaleto{(z)\mathstrut}{4pt}},\\
\end{array}
\end{equation}

where, for each $t$ we have a $q \times 1$ vector $\bm{\nu}_{\scaleto{[i,j,k]t\mathstrut}{4pt}}^{\scaleto{'(z)\mathstrut}{4pt}}$ of observational errors, a $p\times q$ matrix $\Theta_{\scaleto{[i,j,k]t\mathstrut}{4pt}}$ of state parameters, a $p\times q$ matrix $\Omega_{\scaleto{[i,j,k]t\mathstrut}{4pt}}^{\scaleto{(z)\mathstrut}{4pt}}$ of evolution errors. The $1 \times p$  and $p \times p$ matrices $\mathbf{F}^{'}_t$ and $\mathbf{G}_t$ respectivelly are common to each of the $q$ univariate DLMs. The covariates related to the design being used, either a block or an event-related design as well as other characteristics of the subjects, can be included in the columns of $\mathbf{F}^{'}_t$.
For individual analysis the model (\ref{sec2:equ1}) is fitted to every cluster of voxels related to position $[i,j,k]$ in the brain image and the cluster size depends on the $r$ distance, which is a parameter defined by the user. To identify wether or not there is significant evidence of cluster activation or brain reaction at region $[i,j,k]$, three different algorithms (FEST, FSTS and FFBS) proposed in \citet{jimnez2019assessing} are implemented. Those algorithms are sampling schemes that allow to draw on-line trajectory curves related to the state parameter $\Theta_{\scaleto{[i,j,k]t\mathstrut}{4pt}}^{\scaleto{(z)\mathstrut}{4pt}}$, and with those resulting samples a Monte Carlo evidence related to the event of cluster activation is computed. The information obtained from the first stage can be combined in different ways to produce several measures of evidence for the group activation. For a better understanding of the method implemented in this package for the group stage, see \citet{jimnez2019assessing2}. In table \ref{tab:1}, the package's functions for (group) individual analysis and a brief description of them are presented.

\section{Illustrations} \label{sec:illustrations}

\begin{figure}[H]
\centering
\includegraphics[width=.90\textwidth]{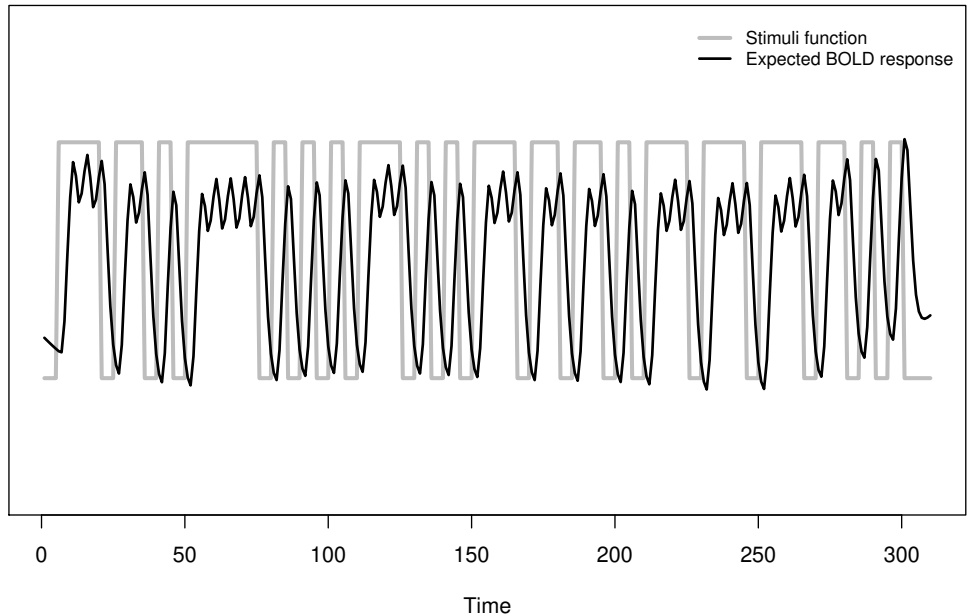}
\caption{\label{fig:1} Merged stimuli function for voice and non-voice sounds with its respective expected BOLD response function for the "voice localizer" example.}
\end{figure}

In order to show some practical illustrations about the use of the \pkg{BayesDLMfMRI} package, we use data related to an fMRI experiment where a sound stimilus is presented. That experiment is intended to offer a "voice localizer" scan, which allows rapid and reliable localization of the voice-sensitive "temporal voice areas" (TVA) of human auditory cortex \citep{pernet2015human}. The data of this "voice localizer" scan is freely available on the online platform OpenNEURO \citep{gorgolewski2017openneuro}. In the original experiment, the voice and non-voice sounds are separately analyzed, but here we merge both sounds in one block as it were just one stimulus (see figure \ref{fig:1}). For the individual analysis, we select one from the 217 subjects whose data are available on OpenNEURO, specifically, we take the data from sub-007. To illustrate the use of the group analysis functions, we take 20 subjects (sub-001:sub-021). The raw fMRI data is preprocessed using the standard processing pipelines implemented on the \pkg{FSL} software for motion correction, spatial smoothing and other necessary procedures to perform statistical analysis for both individual and group stages. 

\subsection*{Individual analysis}

In order to run any of the functions related to individual analysis, the user must provide two inputs: an array of four dimensions containing the MRI images and a matrix whose columns contain the covariates to model the observed BOLD response. Thus, we read the data.nii.qz file, which contains the MRI images, using the function \code{readNIFTI} from the package \pkg{oro.nifti} and the covariates.cvs file which contains the expected BOLD response (show in figure \ref{fig:1}) and its derivative respectively.

\begin{example}
R> library("oro.nifti")
R> fMRI.data <- readNIfTI("./fMRIData/sub-007.nii.gz", reorient=FALSE)
R> fMRI.data <- fMRI.data@.Data
R> dim(fMRI.data)
\end{example}

\begin{example}
[1]  91 109  91 310
\end{example}

\begin{example}
R> Covariates <- read.csv("./covariates.csv", header=FALSE, sep="")
R> dim(Covariates)
\end{example}

\begin{example}
[1] 310   2
\end{example}

To perform a voxel-wise anaylsis \footnote{The BayesDLMfMRI package depends on \pkg{Rcpp}, \pkg{RcppArmadillo}, \pkg{RcppDist} and \pkg{pbapply} packages.} to obtain a 3D array with measurements of activation evidence for every voxel, the user has three options (\code{ffdEvidenceFEST}, \code{ffdEvidenceFFBS} and \code{ffdEvidenceFSTS}), which at the same time can yield three different types of evidence measurements for voxel activation. Just to ilustrate their use, we run them using the "voice localizer" example.

\begin{example}
R> library(devtools)
R> install_github("JohnatanLAB/BayesDLMfMRI")
R> library(BayesDLMfMRI)
R> res <- ffdEvidenceFEST(ffdc = fMRI.data, covariates = Covariates, 
+ m0 = 0, Cova = 100, delta = 0.95, S0 = 1, n0 = 1, Nsimu1 = 100, Cutpos1 = 30, 
+ r1 = 1, Test = "LTT",  Ncores = 15)
\end{example}

\begin{example}
  |++++++++++++++++++++++++++++++++++++++++++++++++++| 100
\end{example}

The arguments \code{m0}, \code{Cova}, \code{S0} and \code{n0}  are the hyper-parameters related to the joint prior distribution of $(\Theta_{\scaleto{[i,j,k]t\mathstrut}{4pt}}^{\scaleto{(z)\mathstrut}{4pt}}, \Sigma_{\scaleto{[i,j,k]t\mathstrut}{4pt}}^{\scaleto{(z)\mathstrut}{4pt}})$. For this example, we are setting a "vague" prior distribution acording to \cite{quintana1987analysis}, where \code{m0 = 0} define a null matrix with zero values in all its entries and both \code{Cova = 100} and \code{S0 = 1} define diagonal matrices. \code{r1} is the euclidean distance, which defines the size of the cluster of voxels jointly modeled. \code{Test} is the parameter related to the test selected by the user, for which there are two options: \code{"LTT"} and \code{"Joint"}. \code{"Ncores"} is the argument related to the number of cores when the process is executed in parallel. \code{Nsimu1} is the number of simulated on-line trajectories related to the state parameter $\Theta_{\scaleto{[i,j,k]t\mathstrut}{4pt}}^{\scaleto{(z)\mathstrut}{4pt}}$. From our own experience dealing with different sets of fMRI data, we recommend \code{Nsimu1 = 100} as a good number of draws to obtain reliable results. \code{Cutpos1} is the time up from where the on-line trajectories are considered in order to compute the activation evidence and \code{delta} is the value of the discount factor.  For a better understanding about the setting of these two last arguments, see (Archive1).

\begin{example}
R> str(res)
\end{example}
\begin{example}
List of 2
 $ : num [1:91, 1:109, 1:91] 0 0 0 0 0 0 0 0 0 0 ...
 $ : num [1:91, 1:109, 1:91] 0 0 0 0 0 0 0 0 0 0 ...
\end{example}

\begin{example}
R> dim(res[[1]])
\end{example}
\begin{example}
[1]  91 109  91
\end{example}

The output for the \code{ffdEvidenceFEST} function depends on the type of \code{Test} set by the user.  For \code{Test = "LTT"} the function returns a list of the type \code{res[[p]][x, y, z]}, where \code{[[p]]} represents the column position in the \code{covariates} matrix and \code{[x, y, z]} represent the voxel position in the brain image. Thus, for the "voice localizer" example \code{res[[1]]} and \code{res[[2]]} are the 3D arrays related to the evidence for brain activation related to the BOLD response for the auditory stimuli and its derivative respectively. When \code{Test = Joint} the output returned is an array 

\begin{example}
R> library(neurobase)
R> res.auxi <- res[[1]]
R> ffd <- readNIfTI("./standard.nii.gz")
R> Z.visual.c <- nifti(res.auxi, datatype=16)
R> ortho2(ffd, ifelse(Z.visual.c > 0.95, Z.visual.c, NA),
+  col.y = heat.colors(50), ycolorbar = TRUE, ybreaks = seq(0.95, 1, by = 0.001))
\end{example}

The \pkg{neurobase} package is one amongst several options available in R to visualize MRI images. In this example, we use its \code{ortho2()} function in order to plot the evidence activation map. The \code{standar.nii.gz} contains the MNI brain atlas, which is used in this work as a reference space for individual and group analysis. For a better understanding of the use of brain atlas, see \cite{brett2002problem}.

\begin{table}[H]
\begin{figure}[H]
  \centering
\begin{center}
\begin{tabular}{cc}
\includegraphics[width=.30\textwidth]{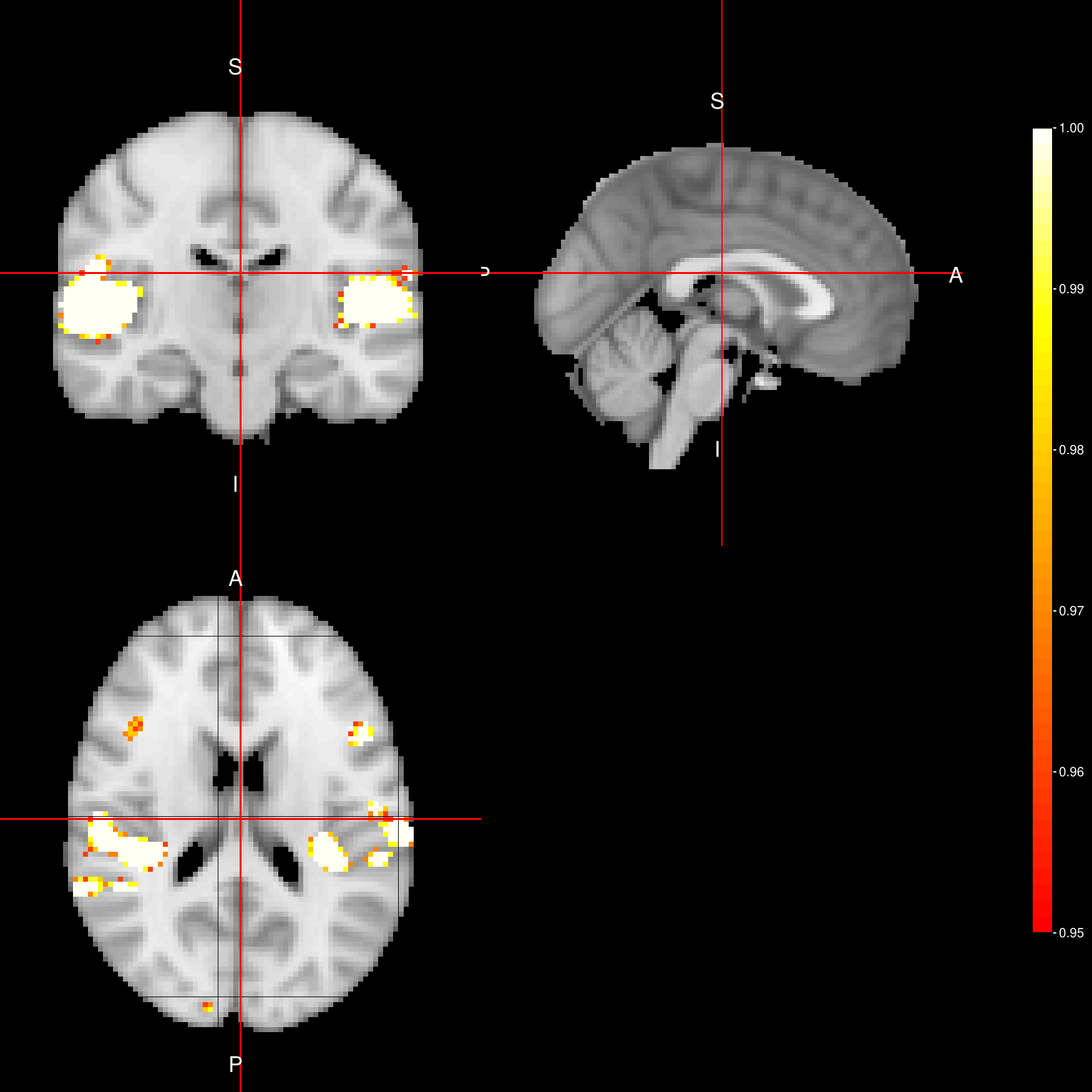}&\includegraphics[width=.30\textwidth]{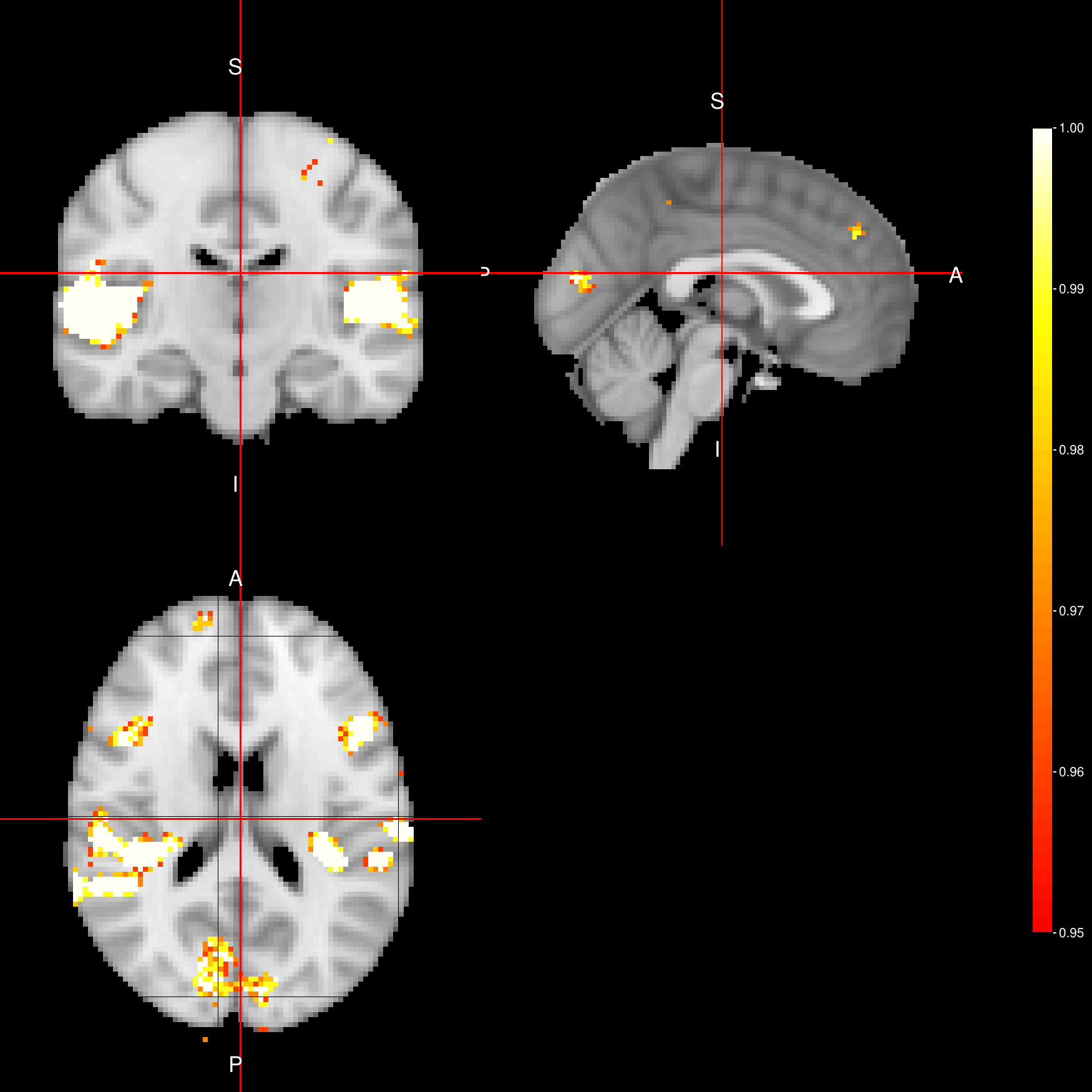}\\
\end{tabular}
\end{center}
  \caption{Activations maps obtained with the functions \code{ffdEvidenceFEST} (left) and \code{ffdEvidenceFFBS} (right) using the "Joint" as sampler distribution.}
  \label{cap2} 
\end{figure}
\end{table}

\begin{example}

R> res <- ffdEvidenceFEST(ffdc = fMRI.data, covariates = Covariates, 
+  m0 = 0, Cova = 100, delta = 0.95, S0 = 1, n0 = 1, Nsimu1 = 100, Cutpos1 = 30, 
+  r1 = 2, Test = "JointTest",  Ncores = 15)
\end{example}

\begin{example}
 |++++++++++++++++++++++++++++++++++++++++++++++++++| 100
\end{example}

\begin{example}
R> str(res)
\end{example}

\begin{example}
List of 4
 $ : num [1:91, 1:109, 1:91] 0 0 0 0 0 0 0 0 0 0 ...
 $ : num [1:91, 1:109, 1:91] 0 0 0 0 0 0 0 0 0 0 ...
 $ : num [1:91, 1:109, 1:91] 0 0 0 0 0 0 0 0 0 0 ...
 $ : num [1:91, 1:109, 1:91] 0 0 0 0 0 0 0 0 0 0 ...\end{example}

For both \code{ffdEvidenceFFBS} and \code{ffdEvidenceFSTS} the input arguments and output structures are the same. Below we run the "voice localizer" example using the \code{ffdEvidenceFFBS} function. It returns a list of the form \code{res[[T]][p,x,y,z]}, where \code{T} defines the type of test (\code{T = 1} for \code{"Marginal"}, \code{T = 2} for \code{"JointTest"}, and \code{T = 3} for \code{"LTT"}),  \code{p} represents the column position in the \code{covariates} matrix and \code{x, y, z} represent the voxel position in the brain image.

\begin{example}
#Change JointTest for Joint
R> res <- ffdEvidenceFFBS(ffdc = fMRI.data, covariates = covariables, m0 = 0, 
+  Cova = 100, delta = 0.95, S0 = 1,n0 = 1, Nsimu1 = 100, Cutpos1 = 30, 
+  r1 = 1,  Ncores = 15)
\end{example}

\begin{example}
 |++++++++++++++++++++++++++++++++++++++++++++++++++| 100
\end{example}

\begin{example}
R> str(res)
\end{example}

\begin{example}
List of 3
 $ : num [1:2, 1:91, 1:109, 1:91] 0 0 0 0 0 0 0 0 0 0 ...
 $ : num [1:2, 1:91, 1:109, 1:91] 0 0 0 0 0 0 0 0 0 0 ...
 $ : num [1:2, 1:91, 1:109, 1:91] 0 0 0 0 0 0 0 0 0 0 ...$
\end{example}

\begin{example}
R> library(neurobase)
R> res.auxi <- res[[3]][1,,,]
R> ffd <- readNIfTI("standard.nii.gz")
R> Z.visual.c <- nifti(res.auxi, datatype=16)
R> ortho2(ffd, ifelse(Z.visual.c > 0.95, Z.visual.c, NA),
+  col.y = heat.colors(50), ycolorbar = TRUE, ybreaks = seq(0.95, 1, by = 0.001))
\end{example}

The \pkg{BayesDLMfMRI} package also has functions that allow taking a closer look for specific voxels defined by the user. For instance, let's suppose we are interested to see some of the output elements related to the FEST algorithm for an active voxel when using the LTT test. 

\begin{example}
R> #Identifying active voxels for a probability threshold of 0.99 
R> active.voxels = which(res[[1]] > 0.99, arr.ind = TRUE)
R> head(active.voxels) 
\end{example}

\begin{example}
     dim1 dim2 dim3
[1,]   28   75   22
[2,]   22   71   24
[3,]   22   72   24
[4,]   23   72   24
[5,]   23   73   24
[6,]   24   73   24
\end{example}

\begin{example} 
R> N1 <- dim(covariables)[1]
R> res.indi <- SingleVoxelFEST(posi.ffd = c(14, 56, 40), covariates 
+  = Covariates, ffdc =  fMRI.data, m0 = 0, Cova = 100, delta = 0.95, S0 = 1, 
+  n0 = 1, Nsimu1 = 100, N1 = N1, Cutpos1 = 30, Min.vol = 0.10, r1 = 1, 
+  Test = "LTT")
R> str(res.indi)  
\end{example}

\begin{example}
List of 4
$ Eviden      : num [1, 1:2] 0.98 0
$ Online_theta: num [1:280, 1:2, 1:100] 2.57 2.52 2.54 2.62 2.46 ...
$ Y_simu      : num [1:280, 1:100] 1.785 0.521 0.624 1.028 -0.165 ...
$ FitnessV    : num 0.762
\end{example}

The function \code{IndividualVoxelFEST()} requires just a few additional input arguments: the position of the  voxel in the brain image (\code{posi.ffd}), the last period of time of the temporal series ($\mathbf{Y}_{\scaleto{[i,j,k]t=1\mathstrut}{5pt}}:\mathbf{Y}_{\scaleto{[i,j,k]t=N1\mathstrut}{5pt}}$) that are used in the analysis (\code{N1}) and \code{Min.vol}, which helps to define a threshold for the voxels that are considered in the analysis. For example, \code{Min.vol = 0.10} means that all the voxels with values below to \code{max(fMRI.data) * Min.vol} are discarted from the analysis. The output is a list containing a vector (\code{Eviden}) with the evidence measure of activation for each one of the \code{p} covariates considered in the model, the simulated online trajectories of $\Theta_{\scaleto{[i,j,k]t\mathstrut}{4pt}}^{\scaleto{(z)\mathstrut}{4pt}}$ (\code{\texttt{Online\_theta}}), the simulated BOLD responses (\code{\texttt{Y\_simu}}) and measure to examine the goodnes of fit of the model ($100 * |\mathbf{Y}_{\scaleto{[i,j,k]t\mathstrut}{5pt}} - \mathbf{\hat{Y}}_{\scaleto{[i,j,k]t\mathstrut}{5pt}}|/\mathbf{\hat{Y}}_{\scaleto{[i,j,k]t\mathstrut}{5pt}}$) for that particular voxel (\code{FitnessV}).

\begin{example} 
R> res.indi2 <- singleVoxelFEST(posi.ffd = c(14, 56, 40), 
+  covariates = covariates, ffdc =  fMRI.data, m0 = 0, Cova = 100,
+  delta = 0.95, S0 = 1,n0 = 1, Nsimu1 = 100, N1=N1, Cutpos1 = 30, Min.vol=0.10,
+  r1 = 1, Test = "JointTest")
R> str(res.indi2)  
\end{example}

\begin{example}
List of 5
$ EvidenMultivariate: num [1, 1:2] 0.81 0
$ EvidenMarginal    : num [1, 1:2] 0.98 0
$ Online_theta      : num [1:2, 1:280, 1:100] 2.38 -1.7 2.42 -1.77 2.35 ...
$ Y_simu            : num [1:100, 1:7, 1:280] 0.597 0.883 1.716 0.11 0.344 ...
$ FitnessV          : num 0.805$
\end{example}

When \code{Test = "Joint"}, the function \code{IndividualVoxelFEST} returns two measures of voxel activation related to the joint and marginal tests along with the already explained remain list elements.

\begin{example} 
R> res.indi3 <- SingleVoxelFFBS(posi.ffd = c(14, 56, 40), covariates = covariates, 
+  ffdc = fMRI.data, m0 = 0, Cova = 100, delta = 0.95, S0 = 1, n0 = 1, 
+  Nsimu1 = 100, N1 = N1, Cutpos1 = 30, Min.vol=0.10, r1 = 1)
R> str(res.indi3)  
\end{example}

\begin{example}
List of 5
 $ Eviden_joint     : num [1, 1:2] 0.94 0.03
 $ Eviden_margin    : num [1, 1:2] 1 0.08
 $ eviden_lt        : num [1, 1:2] 1 0.08
 $ Online_theta     : num [1:310, 1:2, 1:100] 0 0 0 0 0 0 0 0 0 0 ...
 $ Online_theta_mean: num [1:310, 1:2, 1:100] 0 0 0 0 0 0 0 0 0 0 ...$
\end{example}

For both functions \code{IndividualVoxelFFBS} and \code{IndividualVoxelFSTS} the input arguments and output estructures are exactly the same. Three measures of evidence related to the joint, marginal and linear transformation (LTT) tests are generated in the same run. There are also two elements from the list output related to the online simulated trajectories of the state parameter. \code{\texttt{Online\_theta}} are the draws for the marginal model and \code{\texttt{Online\_theta\_mean}} are the mean draws obtained from the joint model.

\begin{example} 
R> frame()
R> plot.window(xlim=c(30, 311), ylim=c(-5, 5))
axis(1, at=seq(30, 310, by = 20), lwd = 2, xlab = "Time")
R> box(lwd = 2)
R> for(i in 1:dim(res.indi$Y_simu)[2]){lines(31:dim(covariates)[1], 
+ res.indi$Y_simu[, i])}
R> lines(31:dim(covariates)[1], covariates[31:dim(covariates)[1], 1],
+ col = "red", lwd = 2)
\end{example}

\begin{example} 
R> frame()
R> plot.window(xlim=c(30, 311), ylim=c(-0.5, 4.5), ylab = expression(theta))
R> axis(1, at=seq(30, 310, by = 20), lwd = 2, xlab = "Time")
R> axis(2, at=-1:4, lwd=2)
R> box(lwd = 2)
R> for(i in 1:dim(res.indi$Online_theta)[2]){lines(31:dim(covariates)[1],
+ res.indi$Online_theta[, i])}
R> abline(h = 0, col = "red" , lwd = 2)
\end{example}

\begin{table}[H]
\begin{figure}[H]
  \centering
\begin{center}
\begin{tabular}{cc}
\code{posi.ffd = c(14, 56, 40) }& \code{posi.ffd = c(28, 67, 15)}\\
\includegraphics[width=.40\textwidth]{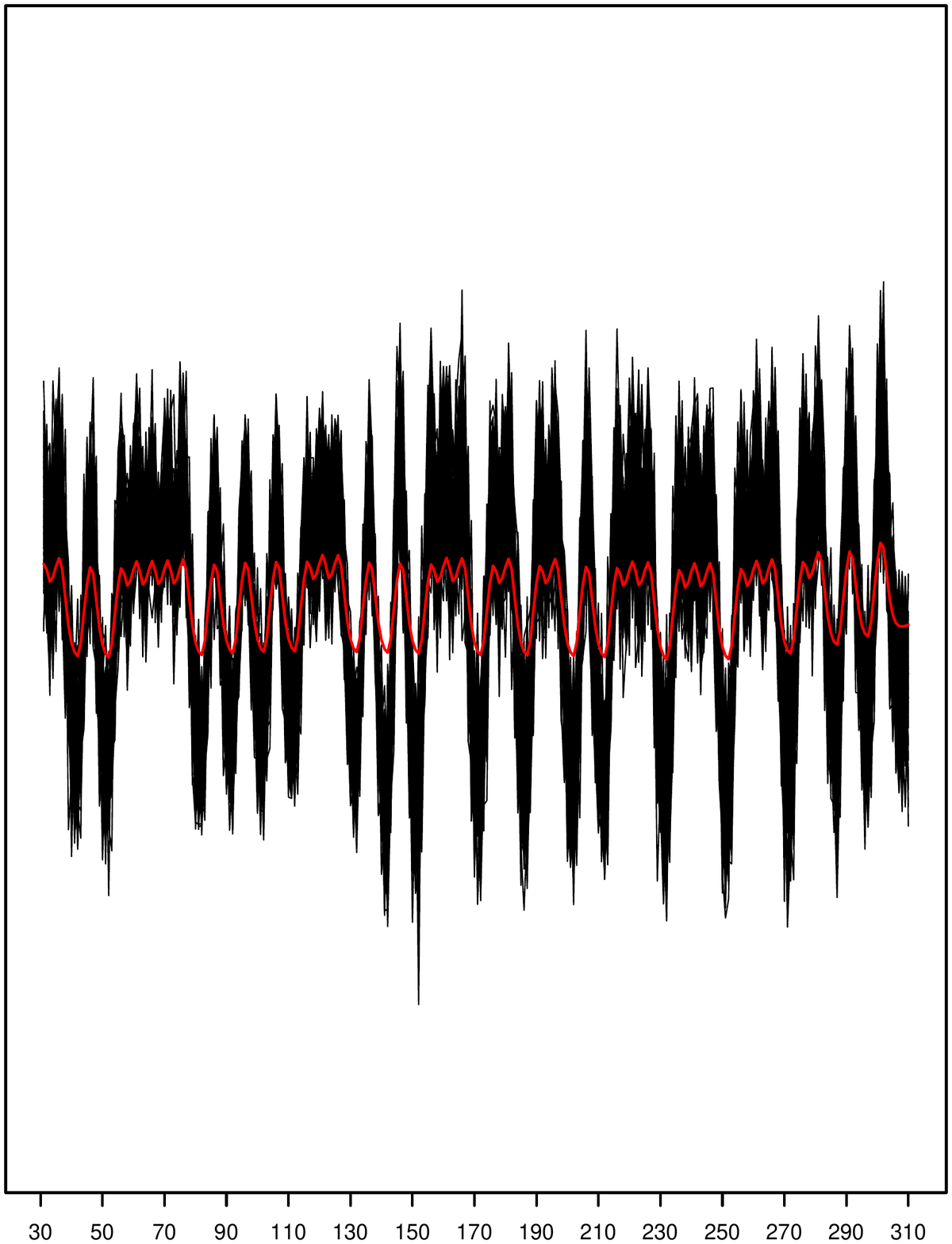}&\includegraphics[width=.40\textwidth]{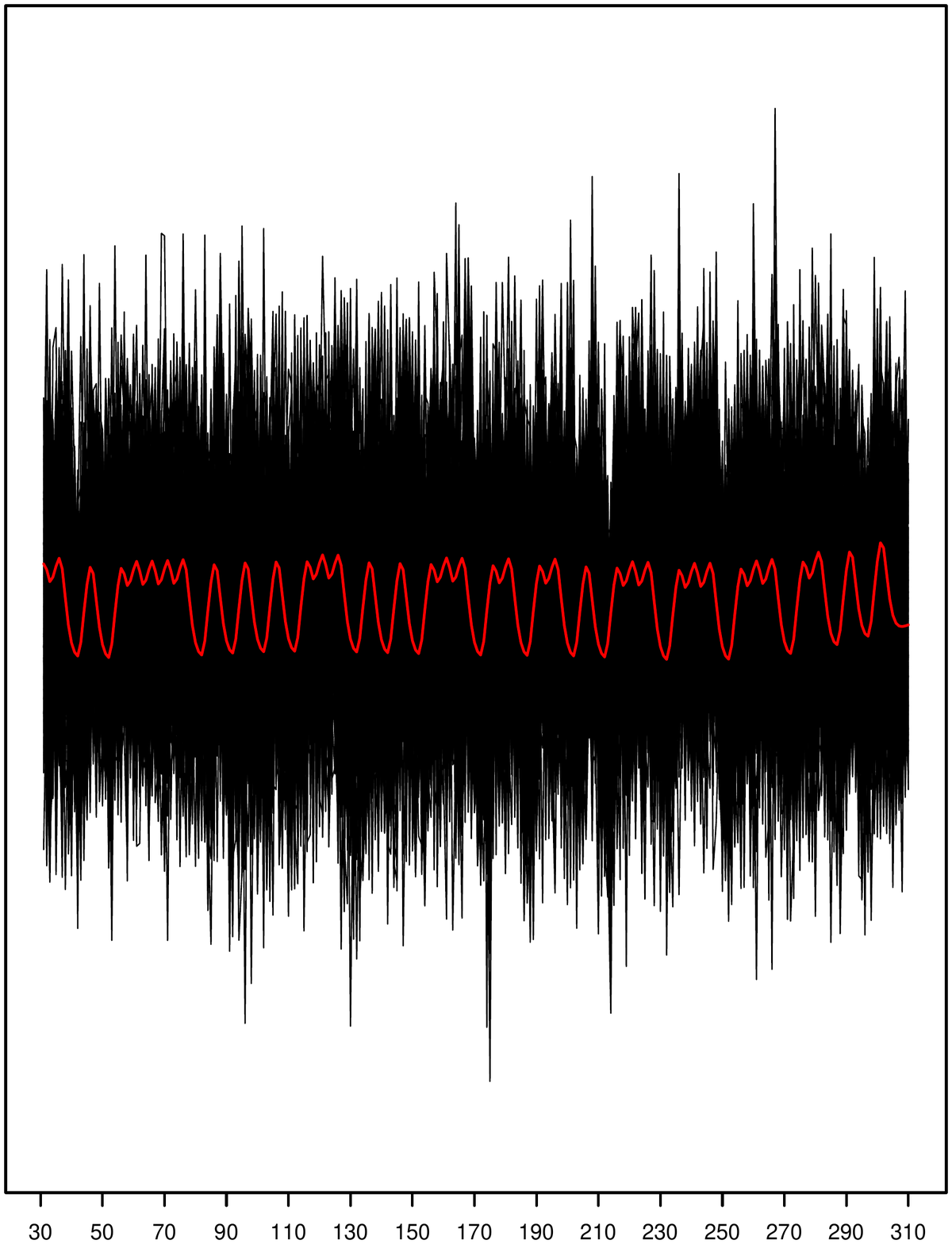}\\
\end{tabular}
\end{center}
  \caption{Simulated Bold response (black lines) and expected Bold response (red line) for an active (left) and non-active (right) voxel respectively.}
  \label{fig3} 
\end{figure}
\end{table}

\begin{table}[H]
\begin{figure}[H]
  \centering
\begin{center}
\begin{tabular}{cc}
\code{posi.ffd = c(14, 56, 40) }& \code{posi.ffd = c(28, 67, 15)}\\
\includegraphics[width=.50\textwidth]{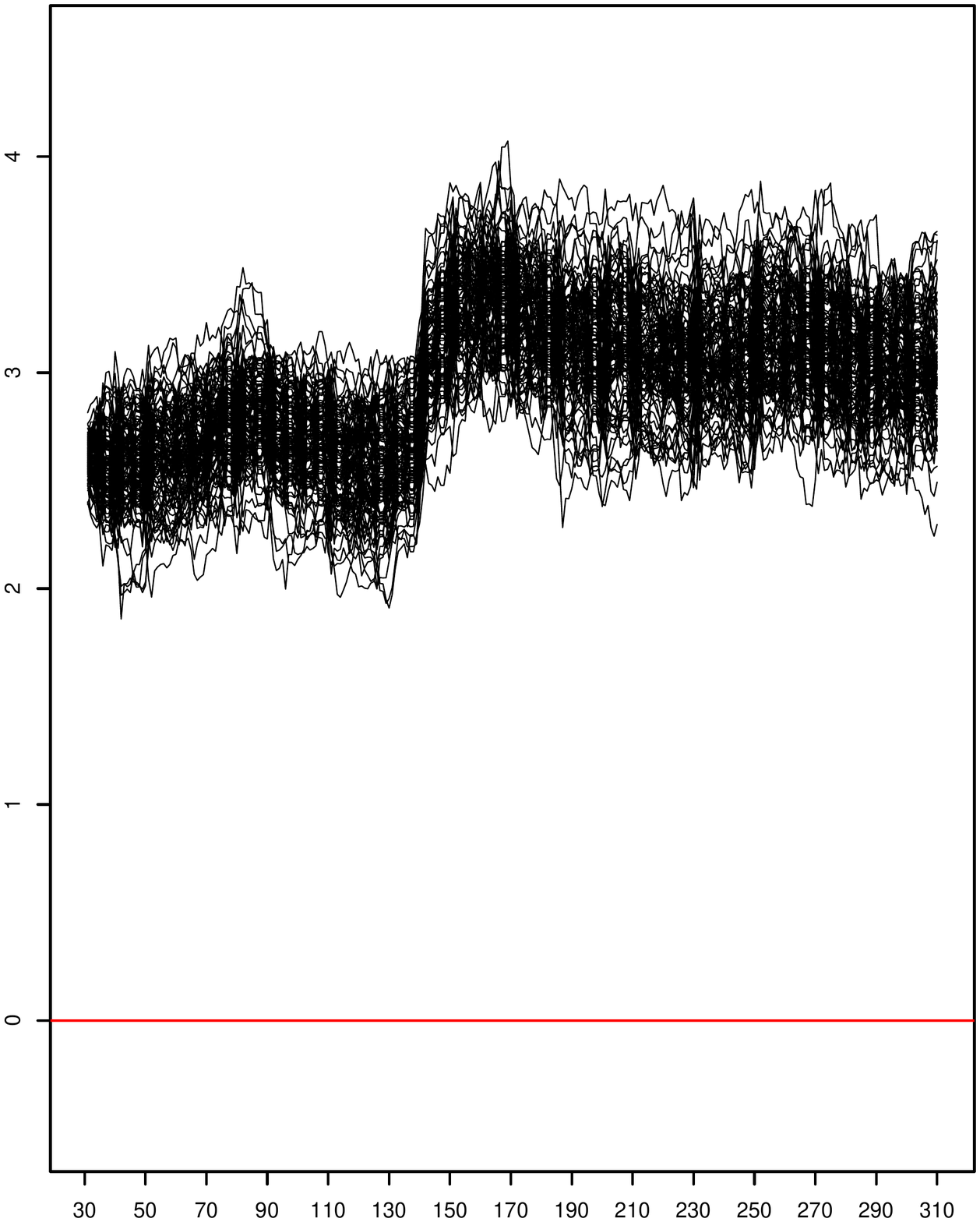}&\includegraphics[width=.50\textwidth]{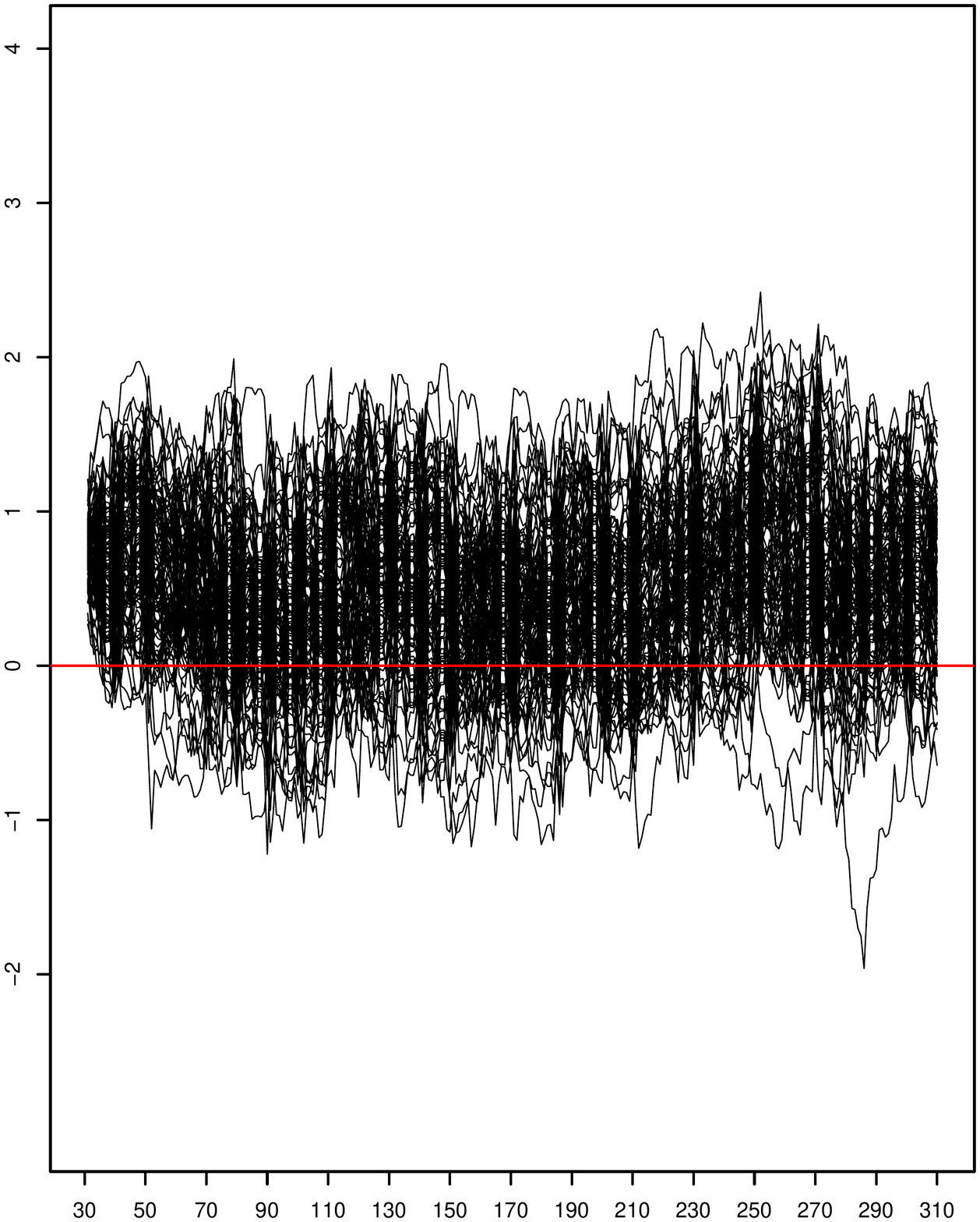}\\
\end{tabular}
\end{center}
  \caption{
  Simulated online trajectories of the state parameter  for an active (left) and nonactive (right) volxel respectively.}
  \label{fig3} 
\end{figure}
\end{table} 

\subsection*{Group analysis}

Now we ilustrate how to run an fMRI group analysis as it is described in (RefARchive2). First, we read the fMRI images of 21 subjects from the "voice-localizer" example:

\begin{example} 
R> names <- list.files("~/fMRIData/")
R> names
\end{example}

using the "Joint" as sampler distribution.

\begin{example}
 [1] "sub-001.nii.gz" "sub-002.nii.gz" "sub-003.nii.gz"
 [4] "sub-004.nii.gz" "sub-005.nii.gz" "sub-006.nii.gz"
 [7] "sub-007.nii.gz" "sub-008.nii.gz" "sub-009.nii.gz"
[10] "sub-010.nii.gz" "sub-011.nii.gz" "sub-012.nii.gz"
[13] "sub-013.nii.gz" "sub-014.nii.gz" "sub-015.nii.gz"
[16] "sub-016.nii.gz" "sub-017.nii.gz" "sub-018.nii.gz"
[19] "sub-019.nii.gz" "sub-020.nii.gz" "sub-021.nii.gz"
\end{example}

\begin{example} 
R> DataGroups <- function(x){
+   ffd.c2 <- readNIfTI(paste("~/fMRIParietal/",x, sep=""), reorient = FALSE) 
+   ffd.c <- ffd.c2@.Data
+   return(ffd.c)
+ }
R> system.time(DatabaseGroup <- parLapply(names, DataGroups, cl = 7))
   user  system elapsed 
637.016 923.008 400.528 
R> str(DatabaseGroup)
\end{example}

\begin{example}
List of 21
 $ : num [1:91, 1:109, 1:91, 1:310] 0 0 0 0 0 0 0 0 0 0 ...
 $ : num [1:91, 1:109, 1:91, 1:310] 0 0 0 0 0 0 0 0 0 0 ...
 $ : num [1:91, 1:109, 1:91, 1:310] 0 0 0 0 0 0 0 0 0 0 ...
 $ : num [1:91, 1:109, 1:91, 1:310] 0 0 0 0 0 0 0 0 0 0 ...
 $ : num [1:91, 1:109, 1:91, 1:310] 0 0 0 0 0 0 0 0 0 0 ...
 $ : num [1:91, 1:109, 1:91, 1:310] 0 0 0 0 0 0 0 0 0 0 ...
 $ : num [1:91, 1:109, 1:91, 1:310] 0 0 0 0 0 0 0 0 0 0 ...
 $ : num [1:91, 1:109, 1:91, 1:310] 0 0 0 0 0 0 0 0 0 0 ...
\end{example}

In order to run any of the functions available in this package to perform fMRI group analysis, the datasets or set of images from each subject must be stored on a list object as it is shown above. To deal with this huge amount of information the user has to have a big RAM memory capacity available on the machine where this process is going to be run. It is also recommended to have a multi-core processor available in order to speed up computation time. The arguments or input parameters for any of the functions offered in this package to run group analysis are almost the same as those required for individual analysis. There is only an additional argument needed  (\code{mask}), which adds a 3D array that works as a brain of reference (MNI atlas) for the group analysis.

\begin{example}
R> MASK <- readNIfTI("~/mask.nii.gz")
\end{example}

\begin{example} 
R> res <- ffdGroupEvidenceFEST(ffdGroup = DatabaseGroup, 
+ covariates = Covariates, m0 = 0, Cova = 100, delta = 0.95, S0 = 1, 
+ n0 = 1, N1 = FALSE, Nsimu1 = 100, Cutpos=30, r1 = 1, Test = "Joint",
+ mask = MASK, Ncores = 7)
\end{example}

\begin{example}
 |++++++++++++++++++++++++++++++++++++++++++++++++++| 100
\end{example}

\begin{example}
R> str(res) 
\end{example}

\begin{example}
List of 4
 $ : num [1:91, 1:109, 1:91] 0 0 0 0 0 0 0 0 0 0 ...
 $ : num [1:91, 1:109, 1:91] 0 0 0 0 0 0 0 0 0 0 ...
 $ : num [1:91, 1:109, 1:91] 0 0 0 0 0 0 0 0 0 0 ...
 $ : num [1:91, 1:109, 1:91] 0 0 0 0 0 0 0 0 0 0 ...
\end{example}

\code{ffdGroupEvidenceFEST} returns an array of dimension $2\times p$ elements, where $p$ is the number of covariates and $2$ is the number of options evaluated as sampler distributions: "Joint" and "Marginal". The first $p$ elements are the 3D arrays related to each column of the \code{covariates} matrix respectively when computing the activation evidence using the "Join" distribution. The ramaining arrays are those related to the marginal distribution.

\begin{example} 
R> res2 <- ffdGroupEvidenceFFBS(ffdGroup = DatabaseGroup, covariates = Covariates, 
+ m0=0, Cova=100, delta = 0.95, S0 = 1, n0 = 1, N1 = FALSE, Nsimu1 = 100, 
+ Cutpos = 30, r1 = 1, mask = MASK, Ncores = 7)
\end{example}

\begin{example}
 |++++++++++++++++++++++++++++++++++++++++++++++++++| 100
\end{example}

\begin{example} 
R> str(res2) 
\end{example}

\begin{example}
List of 3
 $ : num [1:2, 1:91, 1:109, 1:91] 0 0 0 0 0 0 0 0 0 0 ...
 $ : num [1:2, 1:91, 1:109, 1:91] 0 0 0 0 0 0 0 0 0 0 ...
 $ : num [1:2, 1:91, 1:109, 1:91] 0 0 0 0 0 0 0 0 0 0$ ...
\end{example}

\code{ffdGroupEvidenceFFBS} returns an 3D array with the same structure and characteristics as its individual counterpart.

\begin{example}
R> library(neurobase)
R> res.auxi <- res2[[1]][1,,,]
R> ffd <- readNIfTI("standard.nii.gz")
R> Z.visual.c <- nifti(res.auxi, datatype=16)
R> ortho2(ffd, ifelse(Z.visual.c > 0.95, Z.visual.c, NA),
+  col.y = heat.colors(50), ycolorbar = TRUE, ybreaks = seq(0.95, 1, by = 0.001))
\end{example}

\begin{table}[H]
\begin{figure}[H]
  \centering
\begin{center}
\begin{tabular}{cc}
\includegraphics[width=.30\textwidth]{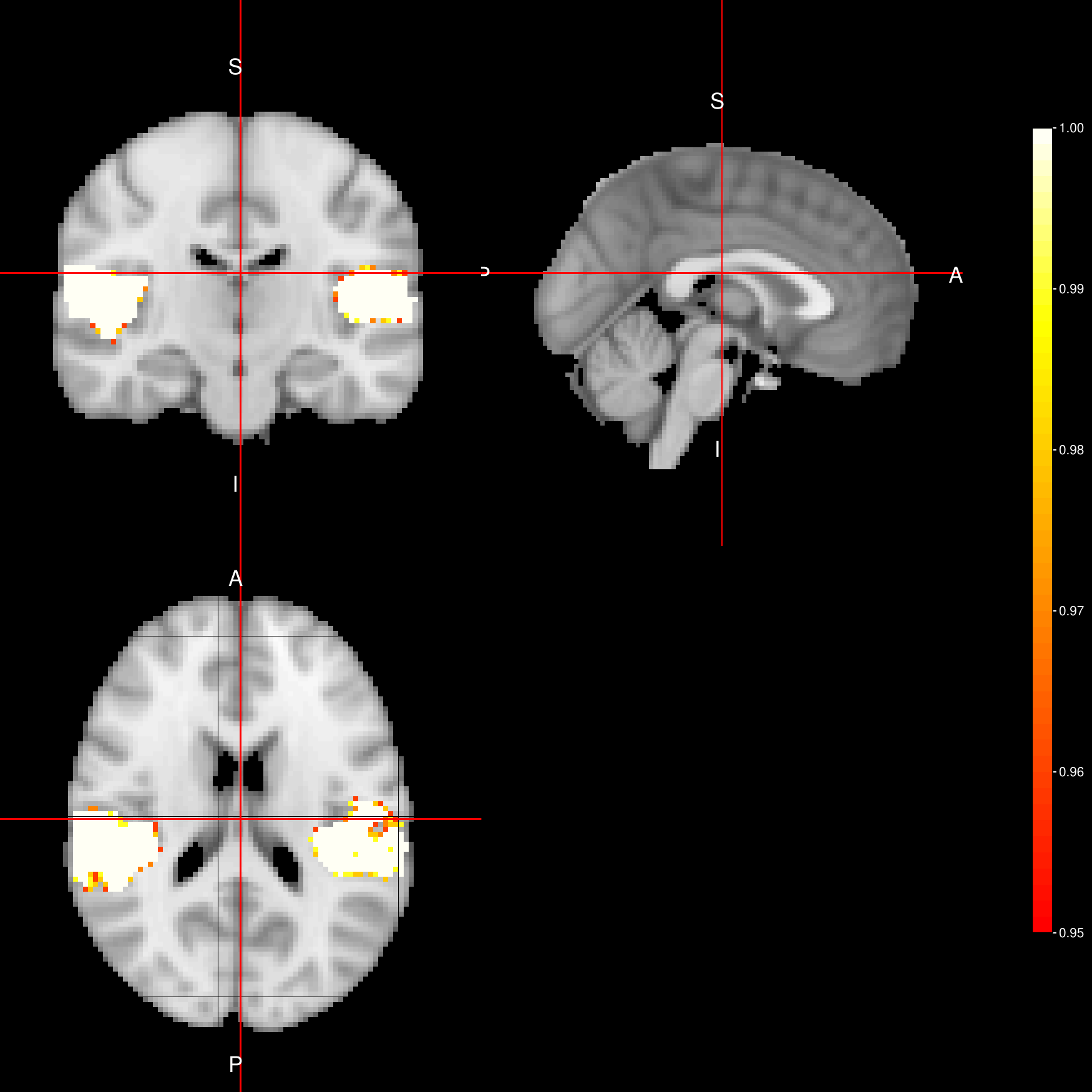}&\includegraphics[width=.30\textwidth]{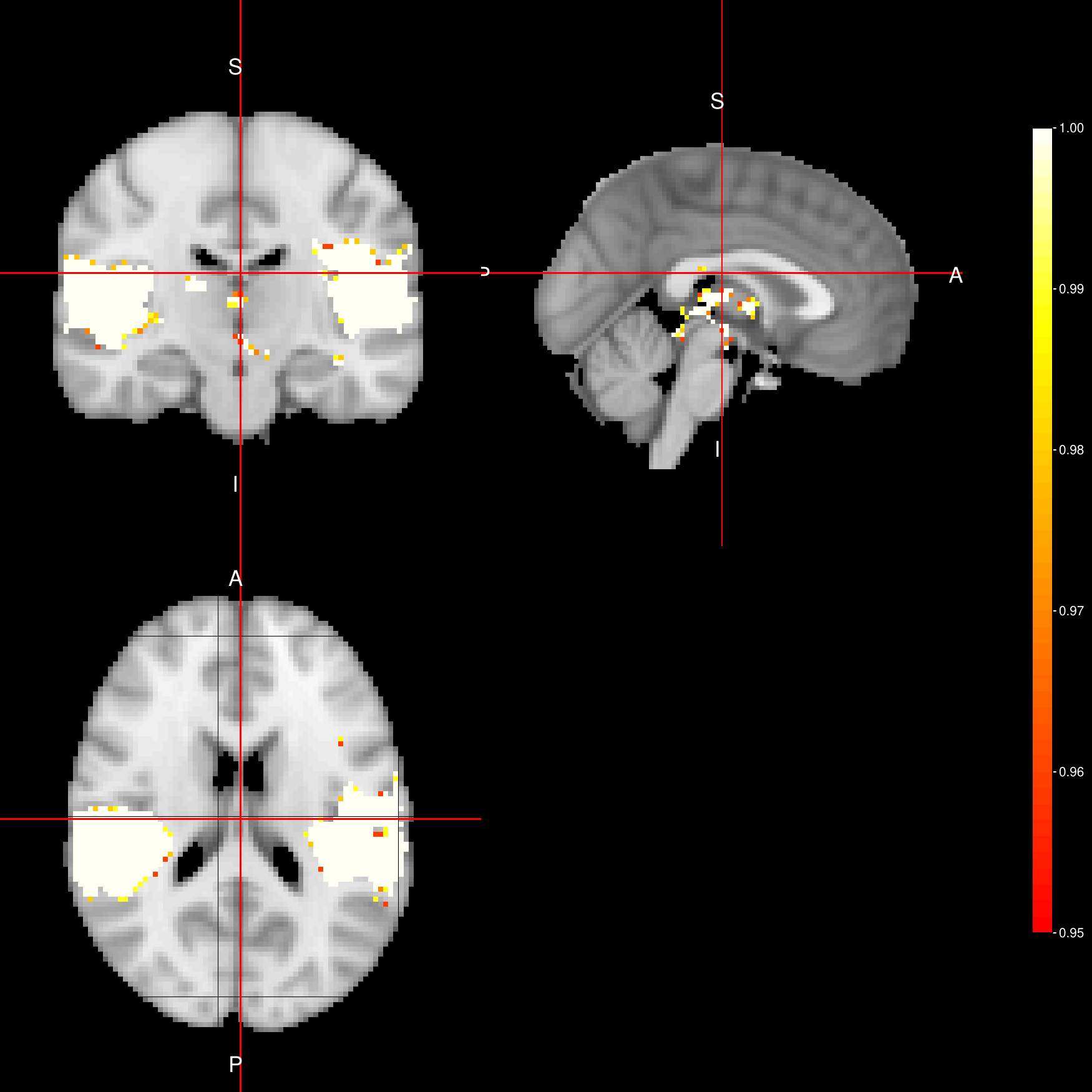}\\
\end{tabular}
\end{center}
  \caption{Activations maps obtained with the functions \code{ffdGroupEvidenceFEST} (left) and \code{ffdGroupEvidenceFFBS} (right) using the "Joint" as sampler distribution.}
  \label{cap2} 
\end{figure}
\end{table} 

The functioning of the functions for single-voxel analysis at the group stage is the same as their counterparts at the individual stage.

\begin{example}
R> resSingle <- GroupSingleVoxelFEST(posi.ffd = c(14, 56, 40), DatabaseGroup,
+ covariates = Covariates, m0 = 0, Cova = 100, delta = 0.95, S0 = 1, n0 = 1, 
+ N1 = FALSE, Nsimu1 = 100, r1 = 1, Test = "Joint", Cutpos = 30)
\end{example}

\begin{example} 
R> frame()
R> plot.window(xlim=c(30, 311), ylim = c(-5, 5))
axis(1, at=seq(30, 310, by = 20), lwd = 2, xlab = "Time")
R> box(lwd = 2)
R> for(j in 2:7){for(i in 1:dim(resSingle[[4]])[1]){lines(31:dim(covariates)[1],
+ resSingle[[4]][i, j,], col = "green")}}
R> for(i in 1:dim(resSingle[[4]])[1]){lines(31:dim(covariates)[1],
+ resSingle[[4]][i, 1,], col = "black")}
lines(31:dim(covariates)[1], covariates[31:dim(covariates)[1], 1], 
+ col = "red", lwd = 2)
\end{example}

\begin{table}[H]
\begin{figure}[H]
  \centering
\begin{center}
\begin{tabular}{cc}
\code{posi.ffd = c(14, 56, 40) }& \code{posi.ffd = c(28, 67, 15)}\\
\includegraphics[width=.40\textwidth]{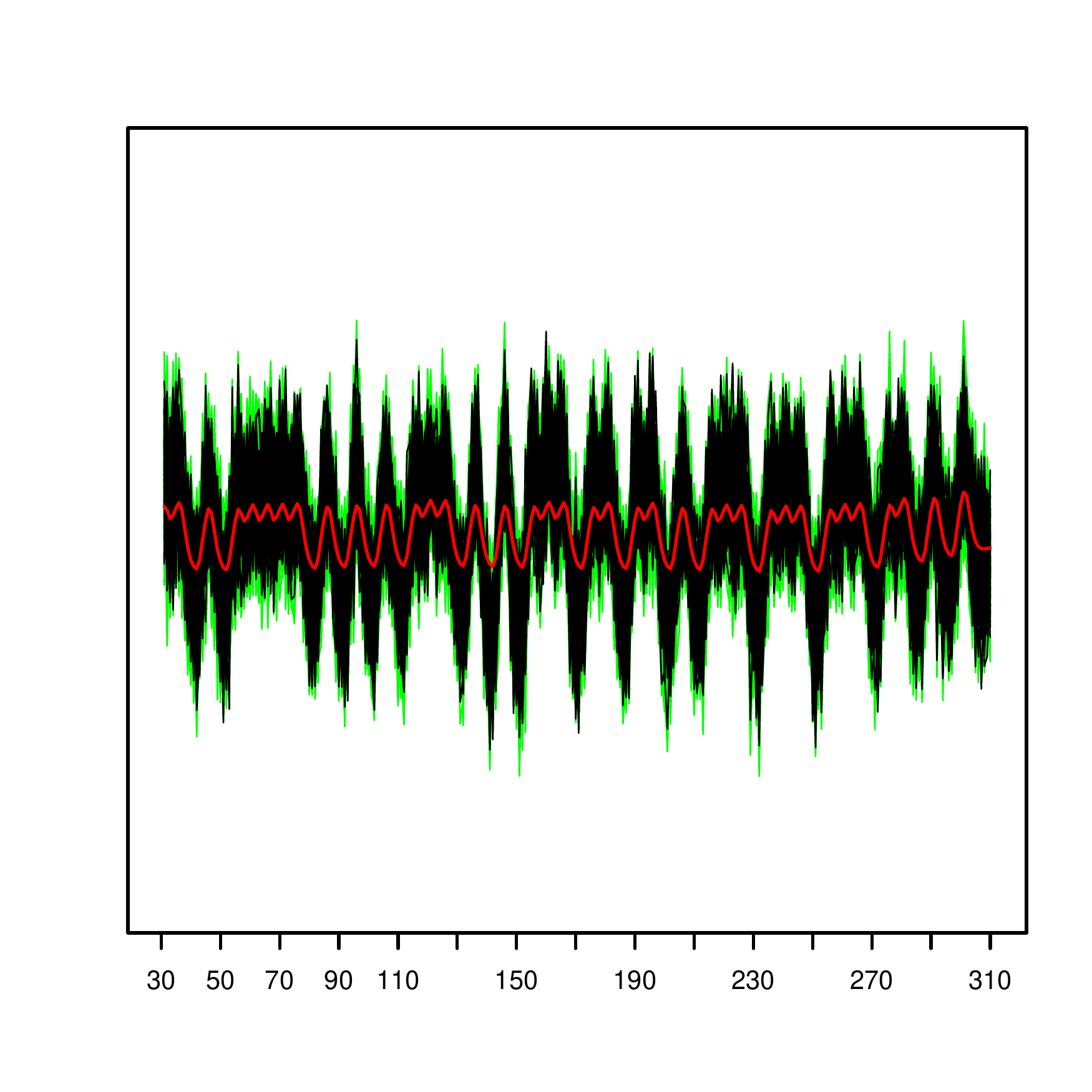}&\includegraphics[width=.40\textwidth]{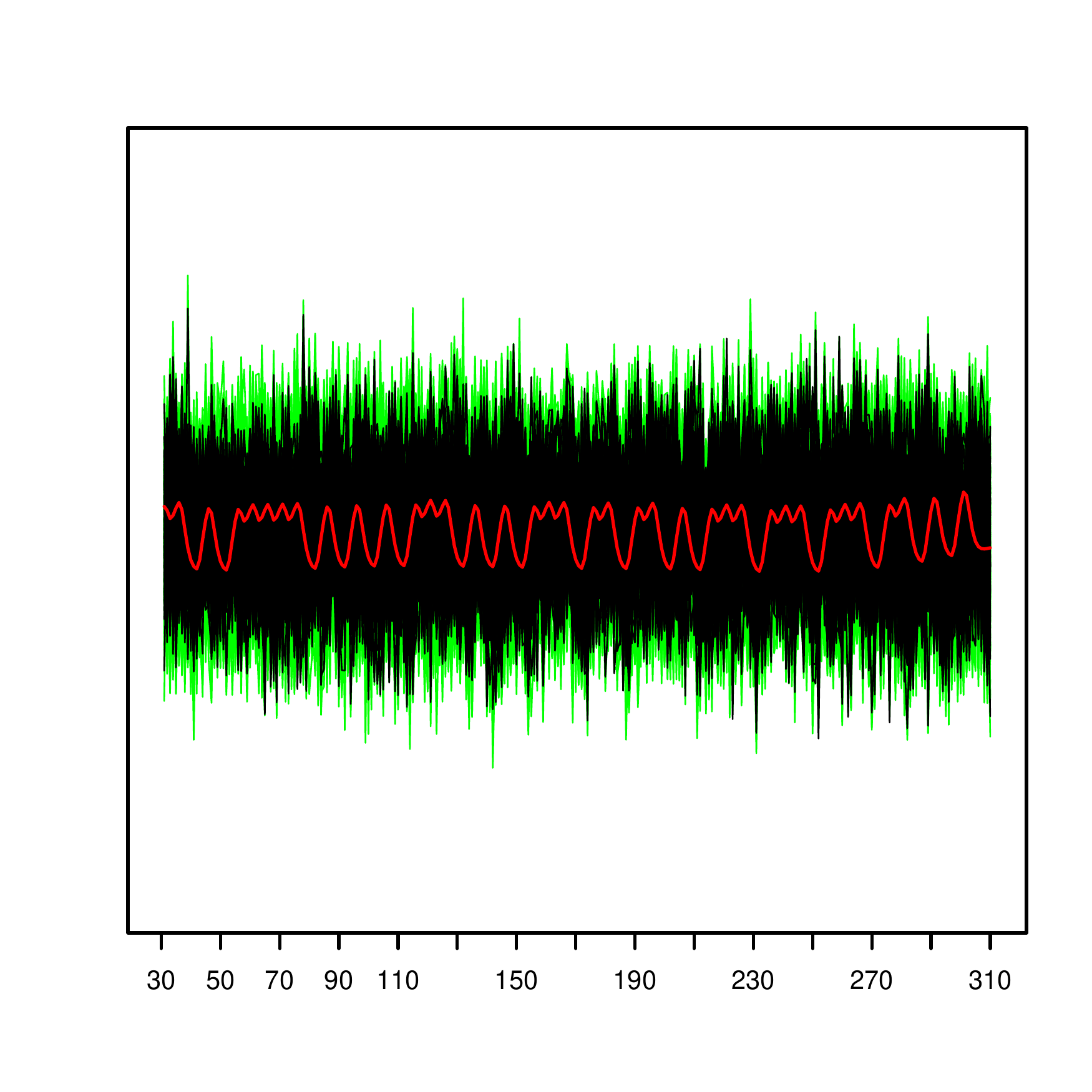}\\
\end{tabular}
\end{center}
  \caption{The black curves are simulated BOLD responses for voxel at position \code{posi.ffd} and the green curves are the BOLD responses related to its neighbors. The red curve is the expected BOLD response for the "voice localizer" experiment.}
  \label{fig3} 
\end{figure}
\end{table} 

\begin{table}[H]
\begin{figure}[H]
  \centering
\begin{center}
\begin{tabular}{cc}
\code{posi.ffd = c(14, 56, 40) }& \code{posi.ffd = c(28, 67, 15)}\\
\includegraphics[width=.50\textwidth]{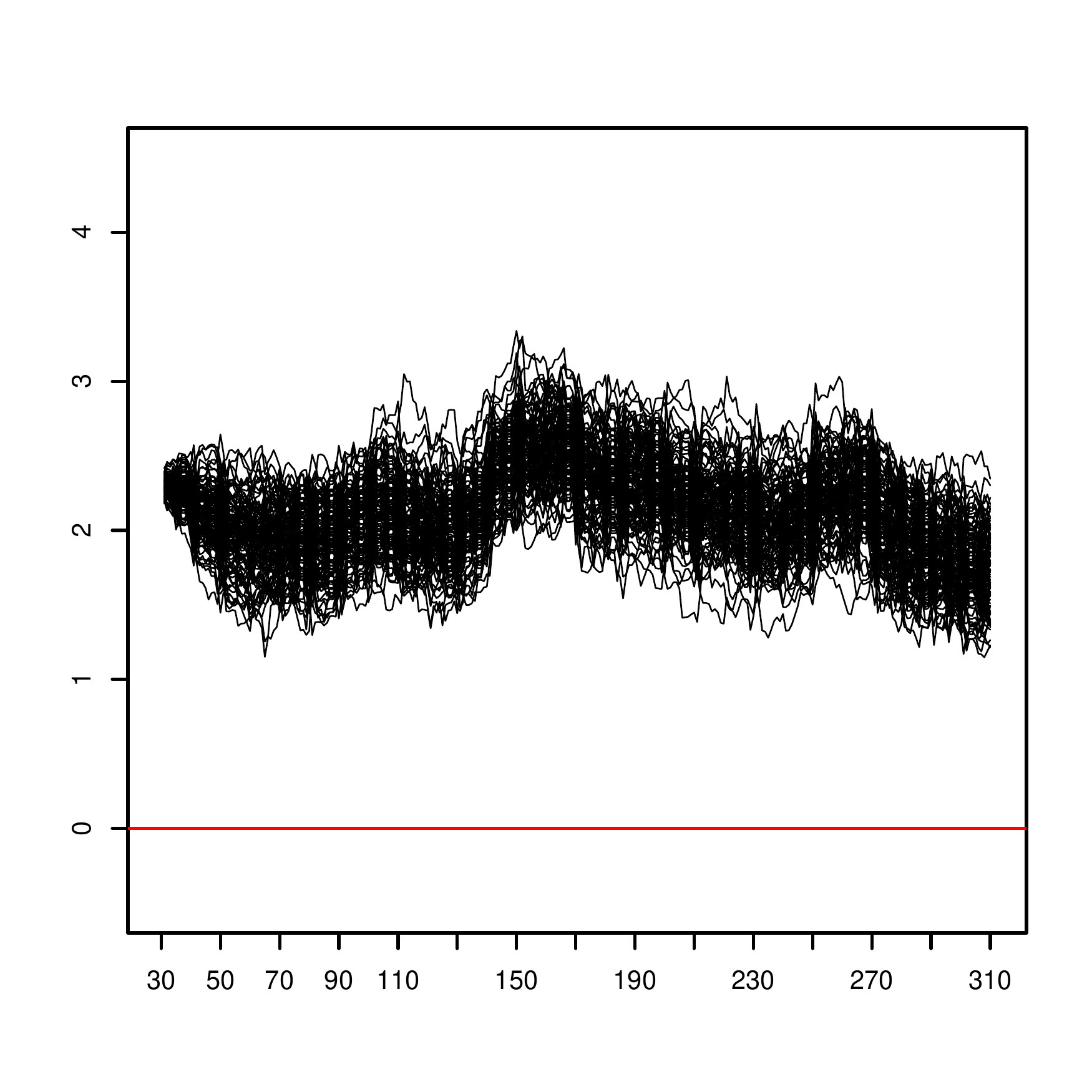}&\includegraphics[width=.50\textwidth]{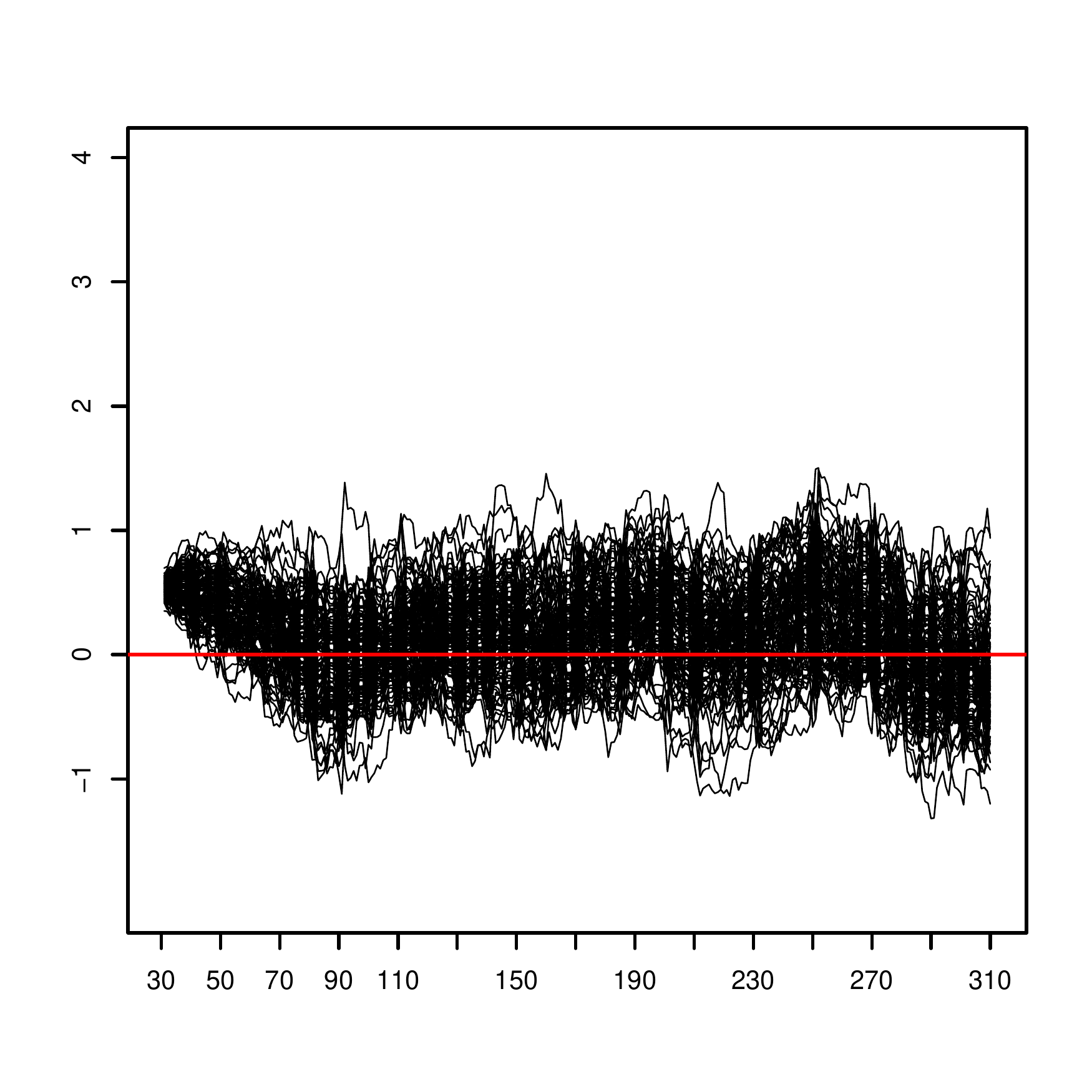}\\
\end{tabular}
\end{center}
  \caption{
  Simulated online trajectories of the state parameter  for an active (left) and nonactive (right) volxel respectively.}
  \label{fig3} 
\end{figure}
\end{table} 

\begin{example} 
R> frame()
R> plot.window(xlim=c(30, 311), ylim=c(-0.5, 4.5), ylab = expression(theta))
R> axis(1, at=seq(30, 310, by = 20), lwd = 2, xlab = "Time")
R> axis(2, at=-1:4, lwd=2)
R> box(lwd = 2)
R> for(i in 1:dim(resSingle[[3]])[3]){lines(31:dim(covariates)[1], 
resSingle[[3]][1, , i])}
R> abline(h = 0, col = "red" , lwd = 2)
\end{example}

\section{Conclusions and future work} \label{sec:summary}

In this work, we present the \pkg{BayesDLMfMRI} package, which allows performing statistical analysis for fMRI data at individual and group stages. It offers different options to assess brain activation for single voxels as well as the entire brain volume and/or more specific brain regions with the help of a mask. The low-level functions are written in C++ and options for parallel computation are available in some of the functions. Currently, some extensions for this package related to comparisons between groups and comparisons between tasks as well as other relevant features are being developed. 

\section*{Computational details}

The results in this paper were obtained using
\pkg{R}~3.4.4 with the \pkg{Rcpp}~1.0.2, \pkg{RcppArmadillo}~0.9.200.5.0 and \pkg{pbapply}~1.4.2 packages on a computer with Linux-Ubuntu, 32 CPUs and 188GB of RAM. \pkg{R} itself
and all packages used are available from the Comprehensive
\pkg{R} Archive Network (CRAN) at
\url{https://CRAN.R-project.org/}.

\bibliographystyle{unsrtnat}
\bibliography{template}

\begin{thebibliography}{26}
\providecommand{\natexlab}[1]{#1}
\providecommand{\url}[1]{\texttt{#1}}
\expandafter\ifx\csname urlstyle\endcsname\relax
  \providecommand{\doi}[1]{doi: #1}\else
  \providecommand{\doi}{doi: \begingroup \urlstyle{rm}\Url}\fi

\bibitem[Zhang et~al.(2016)Zhang, Guindani, Versace, Engelmann, Vannucci,
  et~al.]{zhang2016spatiotemporal}
Linlin Zhang, Michele Guindani, Francesco Versace, Jeffrey~M Engelmann, Marina
  Vannucci, et~al.
\newblock A spatiotemporal nonparametric bayesian model of multi-subject fmri
  data.
\newblock \emph{The Annals of Applied Statistics}, 10\penalty0 (2):\penalty0
  638--666, 2016.

\bibitem[Eklund et~al.(2017)Eklund, Lindquist, and Villani]{eklund2017bayesian}
Anders Eklund, Martin~A Lindquist, and Mattias Villani.
\newblock A bayesian heteroscedastic glm with application to fmri data with
  motion spikes.
\newblock \emph{NeuroImage}, 155:\penalty0 354--369, 2017.

\bibitem[Bezener et~al.(2018)Bezener, Hughes, Jones,
  et~al.]{bezener2018bayesian}
Martin Bezener, John Hughes, Galin Jones, et~al.
\newblock Bayesian spatiotemporal modeling using hierarchical spatial priors,
  with applications to functional magnetic resonance imaging.
\newblock \emph{Bayesian Analysis}, 2018.

\bibitem[Yu et~al.(2018)Yu, Prado, Ombao, and Rowe]{yu2018bayesian}
Cheng-Han Yu, Raquel Prado, Hernando Ombao, and Daniel Rowe.
\newblock A bayesian variable selection approach yields improved detection of
  brain activation from complex-valued fmri.
\newblock \emph{Journal of the American Statistical Association}, \penalty0
  (just-accepted):\penalty0 1--61, 2018.

\bibitem[Zhang et~al.(2015)Zhang, Guindani, and Vannucci]{zhang2015bayesian}
Linlin Zhang, Michele Guindani, and Marina Vannucci.
\newblock Bayesian models for fmri data analysis.
\newblock \emph{Wiley interdisciplinary reviews. Computational statistics},
  7\penalty0 (1):\penalty0 21--41, 2015.

\bibitem[Jenkinson et~al.(2012)Jenkinson, Beckmann, Behrens, Woolrich, and
  Smith]{jenkinson2012fsl}
Mark Jenkinson, Christian~F Beckmann, Timothy~EJ Behrens, Mark~W Woolrich, and
  Stephen~M Smith.
\newblock Fsl.
\newblock \emph{Neuroimage}, 62\penalty0 (2):\penalty0 782--790, 2012.

\bibitem[Penny et~al.(2011)Penny, Friston, Ashburner, Kiebel, and
  Nichols]{penny2011statistical}
William~D Penny, Karl~J Friston, John~T Ashburner, Stefan~J Kiebel, and
  Thomas~E Nichols.
\newblock \emph{Statistical parametric mapping: the analysis of functional
  brain images}.
\newblock Elsevier, 2011.

\bibitem[Kook et~al.(2019)Kook, Guindani, Zhang, and Vannucci]{kook2019npbayes}
Jeong~Hwan Kook, Michele Guindani, Linlin Zhang, and Marina Vannucci.
\newblock Npbayes-fmri: Non-parametric bayesian general linear models for
  single-and multi-subject fmri data.
\newblock \emph{Statistics in Biosciences}, 11\penalty0 (1):\penalty0 3--21,
  2019.

\bibitem[{R Core Team}(2018)]{Rcite}
{R Core Team}.
\newblock \emph{R: A Language and Environment for Statistical Computing}.
\newblock R Foundation for Statistical Computing, Vienna, Austria, 2018.
\newblock URL \url{https://www.R-project.org/}.

\bibitem[Tabelow and Polzehl(2011)]{fmriPackage}
Karsten Tabelow and J\"org Polzehl.
\newblock Statistical parametric maps for functional mri experiments in {R}:
  The package {fmri}.
\newblock \emph{Journal of Statistical Software}, 44\penalty0 (11):\penalty0
  1--21, 2011.
\newblock URL \url{http://www.jstatsoft.org/v44/i11/}.

\bibitem[Whitcher et~al.(2011)Whitcher, Schmid, and Thornton]{oro.nifti}
Brandon Whitcher, Volker~J. Schmid, and Andrew Thornton.
\newblock Working with the {DICOM} and {NIfTI} data standards in {R}.
\newblock \emph{Journal of Statistical Software}, 44\penalty0 (6):\penalty0
  1--28, 2011.
\newblock URL \url{http://www.jstatsoft.org/v44/i06/}.

\bibitem[Muschelli(2018)]{neurobase}
John Muschelli.
\newblock \emph{neurobase: 'Neuroconductor' Base Package with Helper Functions
  for 'nifti' Objects}, 2018.
\newblock URL \url{https://CRAN.R-project.org/package=neurobase}.
\newblock R package version 1.27.6.

\bibitem[Welvaert et~al.(2011)Welvaert, Durnez, Moerkerke, Verdoolaege, and
  Rosseel]{welvaert2011neurosim}
Marijke Welvaert, Joke Durnez, Beatrijs Moerkerke, Geert Verdoolaege, and Yves
  Rosseel.
\newblock neurosim: An r package for generating fmri data.
\newblock \emph{Journal of Statistical Software}, 44\penalty0 (10):\penalty0
  1--18, 2011.

\bibitem[da~Silva et~al.(2011)]{da2011cudabayesreg}
AR~Ferreira da~Silva et~al.
\newblock cudabayesreg: parallel implementation of a bayesian multilevel model
  for fmri data analysis.
\newblock \emph{Journal of Statistical Software}, 44\penalty0 (4):\penalty0
  1--24, 2011.

\bibitem[Sanyal and Ferreira(2019)]{BHMSMAfMRI}
Nilotpal Sanyal and Marco~A.R. Ferreira.
\newblock \emph{BHMSMAfMRI: Bayesian Hierarchical Multi-Subject Multiscale
  Analysis of Functional MRI Data}, 2019.
\newblock URL \url{https://CRAN.R-project.org/package=BHMSMAfMRI}.
\newblock R package version 1.3.

\bibitem[Jim\'enez et~al.(2019)Jim\'enez, de~B.~Pereira, and
  Fossaluza]{jimnez2019assessing}
Johnatan~Cardona Jim\'enez, Carlos~A. de~B.~Pereira, and Victor Fossaluza.
\newblock Assessing dynamic effects on a bayesian matrix-variate dynamic linear
  model: an application to fmri data analysis.
\newblock \emph{arXiv:1910.12058}, 2019.

\bibitem[Jim\'enez(2019)]{jimnez2019assessing2}
Johnatan~Cardona Jim\'enez.
\newblock fmri group analysis based on outputs from matrix-variate dynamic
  linear models.
\newblock \emph{arXiv:1911.00708}, 2019.

\bibitem[Quintana(1985)]{quintana1985}
Jose~Mario Quintana.
\newblock A dynamic linear matrix--variate regression model, 1985.

\bibitem[Eddelbuettel et~al.(2011)Eddelbuettel, Fran{\c{c}}ois, Allaire, Ushey,
  Kou, Russel, Chambers, and Bates]{eddelbuettel2011rcpp}
Dirk Eddelbuettel, Romain Fran{\c{c}}ois, J~Allaire, Kevin Ushey, Qiang Kou,
  N~Russel, John Chambers, and D~Bates.
\newblock Rcpp: Seamless r and c++ integration.
\newblock \emph{Journal of Statistical Software}, 40\penalty0 (8):\penalty0
  1--18, 2011.

\bibitem[Eddelbuettel and Sanderson(2014)]{RcppArma}
Dirk Eddelbuettel and Conrad Sanderson.
\newblock Rcpparmadillo: Accelerating r with high-performance c++ linear
  algebra.
\newblock \emph{Computational Statistics and Data Analysis}, 71:\penalty0
  1054--1063, March 2014.
\newblock URL \url{http://dx.doi.org/10.1016/j.csda.2013.02.005}.

\bibitem[Solymos and Zawadzki(2019)]{pbapplyref}
Peter Solymos and Zygmunt Zawadzki.
\newblock \emph{pbapply: Adding Progress Bar to '*apply' Functions}, 2019.
\newblock URL \url{https://CRAN.R-project.org/package=pbapply}.
\newblock R package version 1.4-1.

\bibitem[Quintana(1987)]{quintana1987multivariate}
Jose~Mario Quintana.
\newblock \emph{Multivariate Bayesian forecasting models}.
\newblock PhD thesis, University of Warwick, 1987.

\bibitem[Pernet et~al.(2015)Pernet, McAleer, Latinus, Gorgolewski, Charest,
  Bestelmeyer, Watson, Fleming, Crabbe, Valdes-Sosa, et~al.]{pernet2015human}
Cyril~R Pernet, Phil McAleer, Marianne Latinus, Krzysztof~J Gorgolewski, Ian
  Charest, Patricia~EG Bestelmeyer, Rebecca~H Watson, David Fleming, Frances
  Crabbe, Mitchell Valdes-Sosa, et~al.
\newblock The human voice areas: Spatial organization and inter-individual
  variability in temporal and extra-temporal cortices.
\newblock \emph{Neuroimage}, 119:\penalty0 164--174, 2015.

\bibitem[Gorgolewski et~al.(2017)Gorgolewski, Esteban, Schaefer, Wandell, and
  Poldrack]{gorgolewski2017openneuro}
Krzysztof Gorgolewski, Oscar Esteban, Gunnar Schaefer, Brian Wandell, and
  Russell Poldrack.
\newblock Openneuro - a free online platform for sharing and analysis of
  neuroimaging data.
\newblock \emph{Organization for Human Brain Mapping. Vancouver, Canada}, 1677,
  2017.

\bibitem[Quintana and West(1987)]{quintana1987analysis}
Jose~Mario Quintana and Mike West.
\newblock An analysis of international exchange rates using multivariate dlm's.
\newblock \emph{The Statistician}, pages 275--281, 1987.

\bibitem[Brett et~al.(2002)Brett, Johnsrude, and Owen]{brett2002problem}
Matthew Brett, Ingrid~S Johnsrude, and Adrian~M Owen.
\newblock The problem of functional localization in the human brain.
\newblock \emph{Nature reviews neuroscience}, 3\penalty0 (3):\penalty0 243,
  2002.

\end{thebibliography}






\end{document}